\documentclass[useAMS,usenatbib]{mn2e}
\usepackage{graphicx}
\usepackage{amsmath}

\def\be{\begin{equation}}
\def\ee{\end{equation}}
\def\ba{\begin{eqnarray}}
\def\ea{\end{eqnarray}}
\def\go{\mathrel{\raise.3ex\hbox{$>$}\mkern-14mu
             \lower0.6ex\hbox{$\sim$}}}
\def\lo{\mathrel{\raise.3ex\hbox{$<$}\mkern-14mu
             \lower0.6ex\hbox{$\sim$}}}

\newcommand{\hatl}{\hat{\mbox{\boldmath $l$}}}
\def\bl{\mbox{\boldmath $l$}}
\def\bomega{{\mbox{\boldmath $\omega$}}}
\def\bmu{{\mbox{\boldmath $\mu$}}}
\def\cN{{\cal N}}
\def\bN{{\bf N}}
\def\br{{\bf r}}
\def\bQ{{\bf Q}}
\def\bcN{{\mbox{\boldmath ${\cal N}$}}}
\def\rin{r_{\rm in}}

\begin{document}
\title[Spin Evolution of Accreting Magnetic Protostars] 
{Evolution of Spin Direction of Accreting Magnetic Protostars 
and Spin-Orbit Misalignment in Exoplanetary Systems}
\author[D.~Lai, F.~Foucart \& D.N.C.~Lin]
{Dong Lai$^{1,2}$\thanks{Email: dong@astro.cornell.edu},
Francois Foucart$^1$, and Douglas N.~C. Lin$^{3,4}$\\
$^1$Center for Space Research, Department of Astronomy, 
Cornell University, Ithaca, NY 14853, USA \\
$^2$Kavli Institute for Theoretical Physics, 
University of California, Santa Barbara, CA 93106, USA\\
$^3$Department of Astronomy and Astrophysics, University
of California, Santa Cruz, CA 95064, USA\\
$^4$Kavli Institute for Astronomy and Astrophysics, Peking University,
Beijing, China\\}

\pagerange{\pageref{firstpage}--\pageref{lastpage}} \pubyear{2010}

\label{firstpage}
\maketitle

\begin{abstract}
Recent observations have shown that in many exoplanetary systems the
spin axis of the parent star is misaligned with the planet's orbital axis.
These have been used to argue against the 
scenario that short-period planets migrated to their present-day
locations due to tidal interactions with their natal discs.  However,
this interpretation is based on the assumption that the spins of young
stars are parallel to the rotation axes of protostellar discs around
them.  We show that the interaction between a magnetic star and its
circumstellar disc can 
(although not always)
have the effect of pushing the stellar spin
axis away from the disc angular momentum axis toward the perpendicular
state and even the retrograde state.  Planets formed in the disc may
therefore have their orbital axes misaligned with the stellar spin
axis, even before any additional planet-planet scatterings or Kozai
interactions take place.  In general, magnetosphere--disc interactions
lead to a broad distribution of the spin--orbit angles, with some
systems aligned and other systems misaligned.
\end{abstract}


\section{Introduction}

\subsection{Background}

Transiting planets are providing new ways to characterize exoplanetary
systems. In particular, the Rossiter-McLaughlin (RM) effect, an
apparent radial velocity anomaly caused by the partial eclipse of a
rotating parent star by its transiting planet, can be used to measure
the sky-projected stellar obliquity, the angle between the stellar
spin axis and the planetary orbital axis. As of August 2010, sky-projected
stellar obliquity has been measured in 28 systems\footnote{The number has 
increased to 48 in October 2010 (A. Triaud, private communication).}
using the RM effect
(see Triaud et al.~2010; Winn et al.~2010).
Among these, about $60\%$ have an orbital
axis aligned (in sky projection) with the stellar spin, while the
other systems show a significant spin-orbit misalignment, including 5 systems 
with retrograde orbits (e.g., H\'ebrard et al.~2008; Winn et al.~2009;
Johnson et al.~2009; Narita et al.~2009; Pont et al.~2010; Jenkins et
al.~2010; Triaud et al.~2010).  In addition, a recent analysis of the
stellar rotation velocity shows that in 10 out of a sample of 75
exoplanetary systems there is likely a significant degree of
misalignment along the line of sight between the planetary orbital
axis and the stellar spin axis (Schlaufman 2010).

It is generally accepted that close-in exoplanets (``hot
Jupiters'') are formed at a distance of order several AU's or larger 
from their host stars before migrating inwards to their current locations.
Gravitational tidal interaction between a gaseous disc and a young planet
(Goldreich \& Tremaine 1980) provides a natural mechanism for
the inward planet migration (Lin et al.~1996; see Lubow \& Ida 2010 for
a recent review). 
Gas-driven migration alone, however, is unlikely to explain the observed 
eccentricity distribution for planets with periods longer than a few weeks.
Two other processes have been suggested to play a role in shaping the 
architecture of exoplanetary systems: (i) Strong gravitational scatterings
between planets in a multiplanet system undergoing dynamical instability
(e.g., Rasio \& Ford 1996; Weidenschilling \& Marzari 1996; Zhou et al.~2007;
Chatterjee et al.~2008); 
(ii) Secular Kozai interactions between a planet and a distant 
companion (a star or planet) in a highly inclined orbit 
(Wu \& Murray 2003; Fabrycky \& Tremaine 2007; Wu et al.~2007; 
see also Eggleton \& Kiseleva-Eggleton 2001)
or between planets in a multiple systems (Nagasawa et al.~2008).
Both scatterings and Kozai interactions are expected to cause
eccentricity in the final planetary orbits and misalignment between the 
stellar spin and the planetary orbit axis.

Newly formed planets are expected to lie in same plane as the gas
disc.  It is usually presumed that the planet-disc interaction
preserves the alignment between the gaseous disc and the orbit of the
planet.  Under the assumption that the stellar spin is aligned with
the disc angular momentum axis, planetary systems with zero or small
stellar obliquity would be produced.  The discovery of a significant
fraction of misaligned systems (particularly the retrograde systems)
has been suggested as a blow against the theory of disc-driven
migration, and to favor the Kozai cycles plus tidal interactions as
the primary mechanism for the formation of hot Jupiters (e.g., Triaud
et al.~2010; Winn et al.~2010).

Planets formed at a few AU's are scattered to the proximity 
of their host stars when their eccentricities approach unity.
Their orbits may be circularized by subsequent tidal dissipation 
in the planets but its efficiency depends sensitively on the
distance of closest approach (e.g., Ivanov \& Papaloizou 2004). Dynamical
relaxation generally leads to a Rayleigh distribution in eccentricity
(Zhou et al.~2007).  If this process is the leading cause for planets
to venture into their stellar proximity, its combined effect with 
the tidal circularization process would yield a continuous periastron 
distribution (Zhang et al. 2010, in preparation), which is not consistent
with the observed sharply bimodal distribution. 

Current data suggest that there may be two populations of 
short-period exoplanet systems, one with spin-orbit alignment, the 
other with significant misalignment (Schlaufman 2010; Winn et al.~2010).
Although stellar or planetary companions have been identified as 
potential culprits for Kozai mechanism to operate in some systems, it 
may not be the dominant process to account for the origin of many hot 
Jupiters which do not appear to be either bound to binary stars or 
associated with planetary siblings with comparable masses and periods 
less than a decade. 

The solar system also provides a clue. Except for Pluto, all planets
outside 1~AU lie within 2$^\circ$ of the ecliptic plane, while the
Sun's equatorial plane is inclined by 7$^\circ$ with respect to the
ecliptic. There is no obvious celestial candidate which can impose
sufficient secular perturbation to induce this observed spin-orbit
inclination.

We note that spin-orbit misalignment is not limited to planetary
systems.  Hale (1994) has measured the inclination to the line of
sight of the spins of stars in binaries by comparing the rotational
period to the $v\sin i$ values of rotationally broadened lines, and
inferred that binaries are spin-aligned for $a\lo 30-40$~AU, but
become randomly oriented for larger orbits. A particularly striking
example is the binary system DI Herculis: Both stars rotate with their
spin axes nearly perpendicular to the orbital axis 
and inclined to each other (Albrecht et al.~2009).
Recently, a number of close binaries (with period of a few days) have 
been found to have nonzero stellar obliquities (A. Triaud, J. Winn, 
private communications, 2010).

\subsection{This Paper}

\begin{figure}
\begin{centering}
\vskip -0.5truecm
\includegraphics[width=9cm]{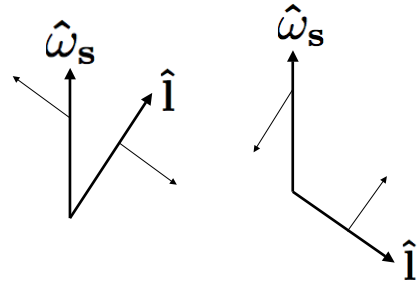}
\vskip -2truecm
\caption{A sketch of the effect of the magnetic warping torque. 
This torque tends to push the disc angular momentum axis $\hat\bl$ toward
a state perpendicular to the spin axis $\hat\bomega_s$. 
When $\hat\bl$ is fixed by the outer disc, the back reaction torque tends 
to push the spin axis toward being misaligned with $\hat\bl$.}
\label{fig1}
\end{centering}
\end{figure}

In this paper and the companion paper (Foucart \& Lai 2010, hereafter
Paper II), we explore an alternative scenario for the observed
spin-orbit misalignment in exoplanet and binary star systems.  We
assume that the planets' present-day orbits are in the plane of their
natal discs.  We study the mutual interaction between a magnetic young
star and its surrounding disc, and show that under certain (but
realistic) conditions, the stellar spin axis can be pushed away from
the disc axis toward the perpendicular state and even the retrograde
state.  Planets formed in the disc therefore may have their orbital
axes misaligned with the stelar spin axis, even before any additional
planet-planet scatterings or Kozai interactions take place. The basic
idea is the following. A magnetic protostar (with $B_\star\go 10^3$~G,
typical for classical T Tauri stars; see, e.g., Bouvier et al.~2007;
Donati \& Landstreet 2009) generally exerts a warping torque and a
precessional torque on the inner region disc before the disc is
disrupted at the magnetosphere boundary (Section 2; see Lai 1999).
These torques have a tendency to make the inner disc tilt away from
the stellar spin axis and precess around it on a warping timescale,
which is much longer than the rotation period
(see Fig.~\ref{fig3} below for a simple ``laboratory'' toy model 
that explains the origin of the warping torque)
\footnote{Note that the secular disc warp discussed in this paper
  is different from the dynamical warp which varies on the timescale
  of the stellar rotation period. Such dynamical warp may arise from
  the periodic vertical force on the disc from the magnetic star
  (Terquem \& Papaloizou 2000; Lai \& Zhang 2008), or from the simple
  effect where the disc material ``climbs'' up the field lines at the
  magnetosphere boundary before funneling to the magnetic polar cap
  [see Bouvier et al.~(2007) for possible observational evidence
  of dynamical warps in the classical T Tauri star AA Tauri].
  The dynamical disc warp averages to zero over an rotation
  period and has no effect on the secular evolution of the system.}.
However, internal processes in the
disc will try to resist the inner disc warping, either by viscous
stress or by bending wave propagation. The result is that, for the
reasonable disc/stellar parameters, the inner disc may not be
significantly warped (see Paper II), i.e., the direction of the inner
disc is approximately aligned with that of the outer disc (assumed to
be fixed; but see below). The back-reaction torque will then act on
the star, changing the spin direction on a timescale much longer than
the disc warping time. Thus, even if initially the stellar spin axis
is approximately (but not perfectly) aligned with the disc axis, given
enough time, the stellar spin axis will evolve towards the
perpendicular state and even the retrograde state (see
Fig.~\ref{fig1}). Therefore, a young planet, formed with its orbit axis aligned
with disc axis, may be misaligned with the stellar spin axis.

While the magnetic warping torque alone pushes the stellar spin towards 
misalignment, other torques (such as that due to the angular momentum
carried by the accreting gas) tend to align the stellar spin with the
disc axis. When these torques are included, the misalignment effect 
is reduced. However, we show that with reasonable disc parameters,
the warping torque could still be dominant and spin-disc misalignment 
can be (but not always) produced under general conditions.

We note that even without the magnetic effect discussed
above, misalignment between stellar spin and disc axis could be 
a natural consequence of star formation process itself. Indeed, 
when stars form under the turbulent conditions of molecular clouds,
the gas accreting onto the protostar does not necessarily 
fall in with a fixed orientation (e.g., McKee \& Ostriker 2007), 
i.e., the outer disc
direction can vary in time. Additionally, the disc can be perturbed 
by other nearby stars and change orientation
(e.g., Bate et al.~2003; Pfalzner et al.~2005).
If the stellar spin is determined by the total angular momentum gained 
over its whole formation history, while the orbit of a planet is 
coplanar with the accretion disc towards the end of its evolution, then
spin-orbit misaligned planetary systems may be produced 
(Bate et al.~2010).
However, since the gas disc carries a large amount of angular momentum,
one might expect that efficient spin-orbit alignment will be achieved
in the absence of any magnetic interaction.
Thus, in this picture, it is important to understand how
the stellar spin evolves and on what timescale (and whether this timescale is
shorter or longer than the timescale of varying outer disc orientation).


The remainder of this paper is organized as follows.  In
Sect.~\ref{sec:analytic}, we present an analytical model used to
describe the interaction between a magnetic star and its disc, and
derive the magnetic torques acting on the disc.  This is similar to
the model considered in Lai (1999), but takes into account of more
recent works on magnetosphere -- disc interactions.  In Sect.~3 we
qualitatively discuss disc warp due to the magnetic torques,
relegating technical details to Paper II.  We then study in Sect.~4
the evolution of stellar spin axis due to the back-reactions of the
magnetic warping torque and other torques, under the assumption that
the disc is approximately flat.
In Sect.~5 we apply our theory to the problem of spin-orbit
misalignment in exoplanetary systems and present two possible
scenarios that may play a role in explaining the observations. Finally
we discuss the implications of our results in Sect.~6.

\section{Analytic Model of Magnetic Star -- Disc Interaction: 
Disc Warping Torque}
\label{sec:analytic}

The interaction between a magnetic star and its circumstellar disc 
has been studied over many decades, both theoretically
(e.g., Pringle \& Rees 1972; 
Ghosh \& Lamb 1979; Aly 1980; 
Wang 1987; Aly \& Kuijpers 1990; 
Shu et al.~1994; 
van Ballegooijen 1994; Lovelace et al.~1995,1999; Lai 1999;
Uzdensky et al.~2002; Pfeiffer \& Lai 2004; D'Angelo \& Spruit 2010)
and using numerical simulations (e.g.,
Hayashi et al.~1996,2000; Miller \& Stone 1997; Goodson et al.~1997; 
Fendt \& Elstner 2000; Matt et al.~2002; Romanova et al.~2003,2009).
Most previous works deal with the case where the stellar spin axis,
the magnetic axis and the disc axis are aligned. Even for such a ``simple'' 
case, the problem is complex. There are at least four physical
ingredients involved in the magnetosphere -- disc interaction: 
(i) The stellar magnetic field penetrates through part of the disc, 
establishing magnetic linkage between the star and the disc. This 
may be achieved either through dissipation in the disc (if the disc is 
sufficiently dissipative) or, more likely, through reconnection between 
the stellar and disc fields -- the latter may be associated with dynamo 
actions in the disc.
(ii) In the inner disc regions where ionization fraction is 
non-negligible, magnetic fields diffuse through the disk on a time scale 
much longer than the orbital period (e.g., Shu et al.~2008; Terquem 2008). 
In this case, the field lines linking the star and the disc are twisted because of the
difference in the stellar rotation $\omega_s$ and disc rotation 
$\Omega(r)$, generating toroidal fields $\Delta B_\phi$ from the vertical
field $B_z$ that threads the disc. After a 
characteristic time of order $|\omega_s-\Omega(r)|^{-1}$, the toroidal
field $|\Delta B_\phi|$ becomes comparable to $|B_z|$, and the flux tube 
starts expanding at a fast rate and the fields open up, temporarily 
disconnecting the star -- disc linkage (e.g., Aly 1985; Aly \& Kuijpers 1990;
van Ballegooijen 1994; Lynden-Bell \& Boily 1994; Lovelace et al.~1995).
However, reconnection between the inflated field lines relaxes the shear
and restore the linkage between the star and the disc, and the whole cycle 
repeats (Aly \& Kuijpers 1990; van Ballegooijen 1994; Goodson et al.~1997;
Matt et al.~2002)\footnote{Strictly steady-state models
have also been considered, but they generally require special conditions
for the disc (see Uzdensky 2004 and references therein).}.
(iii) Some open field lines are strongly ``pinched'' and twisted
by the conducting disc, leading to a conical wind from a localized 
region near the magnetosphere--disc boundary which carries away angular
momentum from the accreting gas (e.g., Shu et al.~1994; 
Romanova et al.~2009).
(iv) In the open field line region near the stellar rotation axis, 
outflows could be launched, carrying away 
angular momentum from the star (e.g., Matt \& Pudritz 2005). Even without
significant mass loss, angular momentum may be transferred outwards 
through Alfven waves.

\begin{figure}
\begin{centering}
\vskip -0.8truecm
\includegraphics[width=10cm]{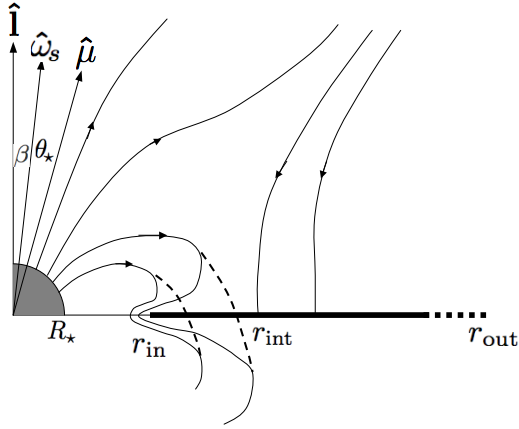}
\vskip -1truecm
\caption{A sketch of magnetic field configuration in a star -- disc
  system for nonzero $\beta$ (the angle between the disc axis and the stellar
  spin axis) and $\theta_\star$ (the angle between the stellar dipole
  axis and the spin axis). Part of the stellar magnetic fields (dashed lines)
  penetrate the disc in the interaction zone between the disc inner
  radius $\rin$ and $r_{\rm int}$ in a cyclic manner, while other field
  lines are screened out of the disc. The closed field lines are
  twisted by the differential rotation between the star and the disc,
  which leads to a magnetic braking torque and a warping torque. The
  screening current in the disc leads to a precessional torque.}
\label{fig2}
\end{centering}
\end{figure}

All these four processes are likely to play a role in the spin
evolution of the protostar. Obviously, for the general case
where the stellar spin axis, magnetic axis and disc axis are misaligned,
the situation is even more complicated. Moreover, the magnetic warping 
and spin-orbit misalignment considered in this paper take place on 
timescales much longer than the dynamical time (i.e., the spin period or
disc orbital period), and therefore cannot be easily 
captured in 3D numerical simulations. 

Nevertheless, the key physical effects of the magnetic star -- disc
interaction relevant to this paper can be described robustly in a
parametrized manner (Lai 1999; see Fig.~\ref{fig2}). The stellar magnetic field
disrupts the accretion flow at the magnetospheric boundary.
Some of the accreting gas are channelled onto the polar caps of the star
while other gas could be ejected in an outflow. The magnetosphere
boundary is located where the magnetic and plasma stresses balance. 
For a dipole field, the inner disc radius is
\be 
r_{\rm in}=\eta \left({\mu^4\over GM_\star\dot M^2}\right)^{1/7},
\label{alfven}
\ee
where $M_\star$ and $\mu$ are the mass and magnetic dipole
moment of the central star,
$\dot M$ is the mass accretion rate, and $\eta$ is a dimensionless constant
of order unity, with typical estimates ranging from
$0.5$ to $1$ and recent numerical simulation giving $\eta\sim 0.5$
(e.g., Long et al.~2005).
Before being disrupted, the disc generally experiences magnetic
torques from the star.
The vertical (perpendicular to the disc) magnetic field produced by 
the stellar dipole is given by
\be
B_z=-{\mu\over r^3}\left(\cos\theta_\star\cos\beta-\sin\theta_\star\sin\beta
\sin\omega_s t\right),
\ee
where $\theta_\star$ is the angle between the magnetic dipole axis 
$\hat\bmu$ and the spin axis $\hat\bomega_s$, and $\beta$ is the
angle between $\hat\bomega_s$ and the disc axis $\hat\bl$ 
(see Fig.~\ref{fig2}) (note that for a warped disc, $\hat\bl$ depends on $r$).
We assume that the static field component, 
$B_z^{(s)}=-(\mu/r^3)\cos\theta_\star\cos\beta$, penetrates the disc
in an ``interaction zone'', between $\rin$ and $r_{\rm int}$.
As discussed before, this field is twisted by the differential rotation
between the star and the disc, undergoing cycles of field inflation 
and reconnection. The differential rotation 
generates toroidal field, whose value at near the disc increases in time until
it becomes comparable to $|B_z^{(s)}|$, at which point reconnection reduces
the field twist, and then the cycle repeats. 
We parameterize the averaged value of the toroidal field generated by the 
twist by 
\be
\Delta B_\phi=\mp\zeta B_z^{(s)},
\label{eq:deltab}\ee
where the upper/lower sign refers to the value above/below the disc plane,
and $|\zeta|\sim 1$
(e.g., Aly 1985; van Ballegooijen 1994; 
Lynden-Bell \& Boily 1994; Lovelace et al. 1995).
The sign of $\zeta$ is such that $\zeta>0$ for
$\Omega(r)-\omega_s\cos\beta>0$, and $\zeta<0$ for
$\Omega(r)-\omega_s\cos\beta<0$. Note that near the corotation radius, 
$\Omega=\omega_s\cos\beta$, the toroidal field may be limited by 
field diffusion inside the disc. Thus, $|\Delta B_\phi/B_z|\sim
|\Omega-\omega_s\cos\beta|\tau\lo 1$ near the corotation radius,
where $\tau$ is the field dissipation timescale for the toroidal field
in the disc. However, such region is very small and can be neglected, 
and we will adopt Eq.~(\ref{eq:deltab}) throughout this paper.
The twisted toroidal field (\ref{eq:deltab}) implies a radial
surface current $K_r=(c/2\pi)\zeta B_z^{(s)}$. The vertical field
$B_z^{(s)}$ acts on $K_r$, giving rise to an azimuthal force
on the disc material and a (well-known) magnetic braking torque
(per unit area)
\be
\bN_{\rm mb}=-{\zeta\over 2\pi}r |B_z^{(s)}|^2\,\hat\bl.
\label{eq:N_mb}\ee
This torque exists even for the aligned case. For $\beta\neq 0$, 
however, there exists an additional torque:
The interaction between $K_r$ and the toroidal 
component of the dipole field, $B_\phi^{(\mu)}=-(\mu/r^3)(\hat\bmu\cdot\hat\phi)$
(where $\hat\phi$ is the unit vector is the azimuthal direction),
gives rise to a vertical force on the disc:
\be
F_z={1\over 8\pi}\left[(B_\phi^{(\mu)}+\zeta B_z^{(s)})^2-
(B_\phi^{(\mu)}-\zeta B_z^{(s)})^2\right]
={\zeta\over 2\pi}B_\phi^{(\mu)}B_z^{(s)}.
\ee

After averaging over the azimuthal angle in the disc and the stellar 
rotation period, the net torque (per unit area) on the disc is
\be
\bN_w=-{\zeta\mu^2\over 4\pi r^5}\cos\beta\cos^2\!\theta_\star
\,{\hat\bl}\times(\hat\bomega_s\times{\hat\bl}).
\label{eq:N_w}\ee
For $\zeta>0$ (i.e., inside the corotation radius),
the effect of this torque (for a fixed spin axis
$\hat\bomega_s$) is to push the disc axis $\hat\bl$ away from
$\hat\bomega_s$ toward the ``perpendicular'' state (see Fig.~\ref{fig1}).
The characteristic warping rate is (for $\beta\sim 0$)
\be
\Gamma_w (r)=\frac{\zeta\mu^2}{4\pi r^7\Omega(r)\Sigma(r)}\cos^2\theta_\star,
\label{eqn:Gamma_w}
\ee
where $\Sigma(r)$ is the surface density of the disc. 
Note that in the case of $\beta=\theta_\star=0$, the corotation radius $r_{\rm co}$
(where $\Omega=\omega_s$) is somewhat larger than $r_{\rm in}$
for stars in the spin equilibrium (e.g., the simulation by Long et al.~2005 
indicates that $r_{\rm co}/r_{\rm in}$ lies in the range of 1.2-1.5).
Thus, for the inner region of the disc most relevant to our paper, 
the condition $\zeta>0$ is satisfied.

The warping torque also exists for more complex stellar magnetic fields,
which may be present for accreting T Tauri stars
(e.g., Hussain et al.~2009; Donati et al.~2010).
Consider an axisymmetric quadrupole field with the symmetric axis
along $\hat\bQ$ (the 3rd axis) such that the magnetic
quadrupole moment tensor satisfies $Q_{11}=Q_{22}=-Q_{33}/2\equiv 
-Q/2$. Outside the star, the stellar quadrupole field is given by
\be
{\bf B}_Q={3Q\over 4r^4}\left[5(\hat\bQ\cdot\hat\br)^2-1\right]\hat\br
-{3Q\over 2r^4}(\hat\bQ\cdot\hat\br)\hat\bQ\,.
\ee
Again, we assume that the static component of the vertical field,
$B_{Q,z}^{(s)}=-(3Q/4r^4)(3\cos^2\theta_Q-1)\sin\beta\cos\beta
\sin\phi$ (where $\theta_Q$ is the angle between $\hat\bomega_s$ and
$\hat\bQ$), penetrates through the disc in a finite region between
$\rin$ and $r_{\rm int}$, and gets twisted to produce 
$\Delta B_\phi=\mp\zeta B_{Q,z}^{(s)}$. After azimuthal averaging
and time averaging over the stellar rotation, we find the warping torque
due to the quadrupole field:
\be
\bN_w^{(Q)}=-{9\zeta Q^2\over 128\pi r^7}(3\cos^2\!\theta_Q-1)^2
\sin^2\!\beta\cos\beta
\,\,{\hat\bl}\times(\hat\bomega_s\times{\hat\bl}).
\ee
Thus, this torque has the same qualitative effect as the warping torque due to
the magnetic dipole. For simplicity, we will neglect $\bN_w^{(Q)}$
in the remainder of this paper.

\begin{figure}
\begin{centering}
\vskip -0.5truecm
\hskip -2truecm
\includegraphics[width=10cm]{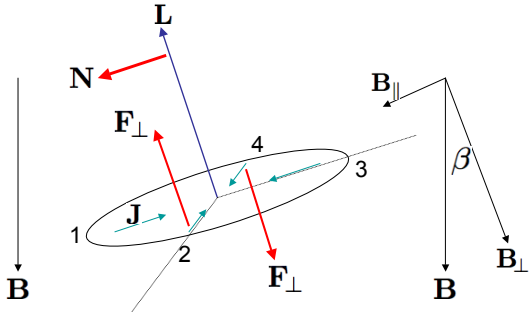}
\vskip -2.1truecm
\caption{A toy model for understanding the origin of the warping
  torque.  A tilted rotating metal plate (with angular momentum 
  ${\bf L}$) in an external magnetic field ${\bf B}$ experiences a
  vertical magnetic force around region 2 and 4 due to the interaction
  between the induced current ${\bf J}$ and the external 
  ${\bf B}_\parallel$, resulting 
  in a torque ${\bf N}$ which further increases the tilt angle
  $\beta$.}
\label{fig3}
\end{centering}
\end{figure}

We note that the existence of the warping torque is rather robust. The
two basic ingredients are: (i) Magnetic field lines from the star
penetrate the inner disc which rotates faster than the star; (ii) the
connected field lines are twisted, reaching a quasi-steady twist
angle. Both ingredients involve dissipations.  Qualitatively speaking,
the warping torque tends to push the system toward the perpendicular
state (with the disc axis perpendicular to stellar spin) in order to
minimize the energy associated with the twisted magnetic fields.  In
fact, the essential physics of the warping torque is captured by the
following ``laboratory'' toy model (cf. Lai 2003), depicted in
Fig.~\ref{fig3}: Consider a metal plate (not a perfect conductor) in
an external uniform magnetic field ${\bf B}$, with the plate surface
initially perpendicular to ${\bf B}$. Neglect gravity. When the plate
rotates (due to some external torque) with the rotation axis ${\bf
  L}$, an EMF is induced (in the direction of ${\bf v}\times {\bf B}$)
and a radial current ${\bf J}$ generated in
the plate -- this is the same surface current $K_r$ mentioned before
eq.~(\ref{eq:N_mb}). 
The interaction between ${\bf J}$ and ${\bf B}$ gives rise
to a ``magnetic braking'' torque which tends to slow down the plate's
rotation.  Now tilt the plate by an angle $\beta$ (and keep the plate
rotating).  The interaction between ${\bf J}$ and ${\bf B}_\parallel$
(the component of ${\bf B}$ parallel to the plate) produces a vertical
force around region 2 and region 4 in the plate. The result is a
torque ${\bf N}$ which tends to push the plate's rotation axis ${\bf
  L}$ away from the magnetic direction, further increasing the tilt.

In addition to the warping torque, there is also a precessional torque
on the disc when $\beta\neq 0$. This arises from the dielectric
property of the disc: If the disc
does not allow the vertical stellar field (e.g., the rapidly varying
component of $B_z$ due to stellar rotation) to penetrate, an azimuthal
screening current $K_\phi$ will be induced in the disc. This $K_\phi$
interacts with the radial magnetic field $B_r$ from the stellar dipole
and produces a vertical force. After azimuthal averaging and 
averaging over the stellar rotation, we obtain the torque per unit area:
\footnote{
This assumes that only the spin-variable vertical field is
screened out by the disc. If the vertical field is entirely 
screened out, then $(-\sin^2\theta_\star)$ should be replaced by 
($2\cos^2\theta_\star-\sin^2\theta_\star$).} 
\be
\bN_p=-\frac{\mu^2}{\pi^2 r^5 D(r)}
\sin^2\theta_\star\cos\beta\,\hat\bomega_s
\times\hat\bl,
\label{eq:N_p}\ee
where the dimensionless function $D(r)$ is given by
\begin{equation}
D(r)={\rm max}~\left(\sqrt{r^2/r^2_{\rm in}-1}, \sqrt{2H(r)/r_{\rm in}}\right),
\label{eqn:D(r)}
\ee
and $H(r)$ is the half-thickness of the disc.
The torque (\ref{eq:N_p}) tends to make the disc precess around the 
stellar spin axis.


\section{Warped Discs}
\label{sec:warp}

As shown in Sect.~2, the inner region of the disc where
magnetic field lines connect star and the disc and where the disc
rotates faster than the star ($\Omega>\Omega_s\cos\beta$)
experiences a warping torque and a precessional torque.
If we imagine dividing the disc into many rings, and if each ring
were allowed to behave independent of each other, it would 
be driven toward a perpendicular state ($\hat\bl \perp\hat\bomega_s$)
and precess around the spin axis
of the central star. Obviously, real protoplanetary discs do not behave
as a collection of non-interacting rings: Hydrodynamic force and viscous
force provide strong couplings between different rings. Depending on the
physical condition of the disc, the dynamics of a warped disc may be driven 
by viscous diffusion (when the viscous $\alpha$-parameter is greater than
the dimensional disc thickness $H/r$) or propagation of bending waves (when
$\alpha\lo H/r$) (Papaloizou \& Pringle 1983; Paper II).

Thus, for a given outer disc direction $\hat\bl_{\rm out}$ and 
stellar spin axis $\hat\bomega_s$, the disc will generally evolve into
a warped state, with $\hatl$ dependent on $r$.  In particular, the
orientation of the disc at the inner radius, $\hatl_{\rm in}
=\hatl(r_{\rm in})$, generally differs from $\hatl_{\rm out}$.  For
typical protostar parameters, $M_\star\sim 1~M_\odot$, $R_\star\sim
2R_\odot$, $\mu=B_\star R_\star^3$ (with the surface field
$B_\star\sim 10^3$~G), and mass accretion rate $\dot M\sim
10^{-8}M_\odot\,{\rm yr}^{-1}$, the inner disc is located at a few
stellar radii. From equation (\ref{eqn:Gamma_w}), we find that
the timescale for the warp evolution is of order 
\ba
&&\Gamma_w^{-1}=\left(92\,{\rm days}\right) \left({1~{\rm kG}\over
  B_\star}\right)^{\!2} \left({2R_\odot\over R_\star}\right)^{\!6}
\left({M_\star\over 1\,M_\odot}\right)^{\!1/2}\nonumber\\ 
&&\qquad
\times \left({r\over 8R_\odot}\right)^{\!11/2} 
\left({\Sigma\over 10\,{\rm g\,cm}^{-2}}\right)
\left(\zeta\cos\theta_\star\right)^{-1}.
\ea 
As shown in Paper II, under the combined actions of 
warping/precessional torques and disc viscosity or bending waves, the
disc will settle down into a steady state on a timescale 
that depends on disc viscosity and sound speed
[see related work by Papaloizou \& Terquem (1995)
for discs in binary stars with inclined orbits.]
In general, this timescale
can be up to several orders of magnitude of $\Gamma_w^{-1}$ evaluated at
$r_{\rm in}$. Nevertheless,
it is much shorter than the disc lifetime and the timescale of
the secular evolution of the stellar spin (see Eq.~\ref{eq:tspin}
below).  Moreover, we show in Paper II that for most reasonable stellar/disc
parameters, the steady-state disc warp is rather small because of
efficient viscous damping or propagation of bending waves.  Thus, we
will adopt the approximation $\hat\bl(r)\simeq \hat\bl_{\rm out}$ in
the remainder of this paper, leaving more detailed study of the effect
of warped discs to Paper II.

\section{Secular Evolution of Stellar Spin}
\label{sec:flat}


We now consider the evolution of the stellar spin direction due to the 
back-reaction of the magnetic warping torque and other torques. 
Note that we are interested
in the long-term evolution of the stellar spin. 
The specific angular momentum of the accreting gas at $r_{\rm in}$ is
given by the Keplerian value $\sqrt{GM_\star r_{\rm in}}$, so 
the characteristic accretion torque on the star is 
\be
\cN_0=\dot M(GM_\star\rin)^{1/2}.
\ee
Thus the fiducial timescale for the spin evolution is
\ba
&&t_{\rm spin}={J_s\over \cN_0}
=(1.25\,{\rm Myr})\left(\!{M_\star\over 1\,M_\odot}\!\right)
\!\!\left({{\dot M}\over 10^{-8}{M_\odot}{\rm yr}^{-1}}\right)^{\!-1}
\nonumber\\
&&\qquad \times \left(\!{\rin\over 4R_\star}\!\right)^{\!-2}
\!\!{\omega_s\over\Omega(\rin)},
\label{eq:tspin}\ea
where we have assumed the stellar spin angular momentum 
$J_s=0.2M_\star R_\star^2 \omega_s$ 
(for a fully convective protostar modeled as a $\Gamma=5/3$ polytrope).

In general, the evolution equation for the spin angular momentum of the star, 
$J_s\hat\bomega_s$, can be written in the form
\be
{d\over dt}\left(J_s\hat\bomega_s\right)=\bcN=
\bcN_l+\bcN_s+\bcN_w+\bcN_p.
\label{spin}\ee
Here $\bcN_l$ represents the torque component that is aligned with the
inner disc axis $\hat\bl_{\rm in}$ ($=\hat\bl$ for a flat disc):
\be
\bcN_l=\lambda\dot M (GM_\star\rin)^{1/2}\,\hatl,
\label{eq:Nl}\ee
where $\lambda$ is a dimensionless parameter. Equation (\ref{eq:Nl})
includes not only the accretion torque carried by the accreting gas onto the star,
$\dot M_{\rm acc} (GM\rin)^{1/2}\hatl$ (note that in general, the
accretion rate onto the star, $\dot M_{\rm acc}$,
may be smaller than $\dot M$, the disc accretion rate), 
but also includes the magnetic braking torque associated with the
disc -- star linkage, as well as any angular momentum carried away by
the wind from the magnetosphere boundary. While the details are complex,
we would expect that all these contributions to $\bcN_l$ tend to 
make $\lambda<1$, as suggested by recent
numerical simulations (e.g., Romanova et al.~2009) and earlier works
(e.g., Shu et al.~1994).
The term $\bcN_s=-|{\cal N}_s| \hat\bomega_s$ represents a spindown torque
carried by a wind/jet from the open field line region of the star.
In the aligned case ($\hat\bl=\hat\bomega_s$) studied by previous works,
$\bcN_l$ and $\bcN_s$ are the only torques acting on the star,
and spin equilibrium is reached when 
$\lambda\dot M (GM_\star\rin)^{1/2}=|\cN_s|$. Note that in general,
the values of $\lambda$ and $\cN_s$ may depend on $\beta$ and other 
quantities.

The term $\bcN_w$ and $\bcN_p$ represent the back-reactions
of the warping and precessional torques, respectively. 
From Eq.~(\ref{eq:N_w}), we have
\ba
&&\bcN_w=-\int_{\rin}^{r_{\rm int}}\! 2\pi r {\bf N}_w\,dr\nonumber\\
&&\qquad ={\zeta'\mu^2\over 6\rin^3}\cos^2\theta_\star\cos\beta\,
\hatl\times (\hat\bomega_s\times\hatl)\nonumber\\
&&\qquad =\cN_0 n_w\,\hatl\times (\hat\bomega_s\times\hatl),
\label{eq:bcN_w}\ea
where
\be
n_w={\zeta'\over 6\eta^{7/2}}\cos^2\!\theta_\star\,\cos\beta,
\ee
with\footnote{This expression for $\zeta'$ assumes that the corotation
radius $r_{\rm co}$ is larger than $r_{\rm int}$.}
\be
\zeta'=\zeta \left[1-(\rin/r_{\rm int})^3\right].
\ee
Similarly, from Eq.~(\ref{eq:N_p}), we have
\be
\bcN_p=-\int_{\rin}^\infty\! 2\pi r {\bf N}_p\,dr
=\cN_0 n_p\,\hat\bomega_s\times\hatl,
\ee
with (for thin discs)
\footnote{The coefficient $4\over3$ in equation (\ref{eq:np}) is only
valid in the limit $\delta=H/r \rightarrow 0$.  For small but finite
$\delta$, we get ${4\over3}[1-{3\over2}\sqrt{\delta \over 2}+O(\delta)]$.},
\be
n_p={4 \over 3\pi\eta^{7/2}}\sin^2\!\theta_\star\,
\cos\beta.\label{eq:np}
\ee
Note that both $\bcN_w$ and $\bcN_p$ are of order $\mu^2/\rin^3$ (see
the first line of Eq.~[\ref{eq:bcN_w}]), which does not directly
depend on $\dot M$.  However, when we use Eq.~(\ref{alfven}) for
$\rin$, we find that both $\bcN_w$ and $\bcN_p$ are of the same order
of magnitude as the fiducial accretion torque $\cN_0$. Therefore, as
we will see below, the timescale to change the stellar spin direction (if
at all) is of the same order as $t_{\rm spin}$.

From Eq.~(\ref{spin}), we find that 
the magnitude of the stellar spin evolves according to
\be
{d\over dt}J_s=\bcN\cdot\hat\bomega_s=\cN_0\left(\lambda\cos\beta
+n_w\sin^2\!\beta\right)-|\cN_s|.
\ee
The inclination angle of the stellar spin relative to the disc
evolves according to the equation
\be
{d\over dt}\cos\beta={\cN_0\over J_s}\sin^2\!\beta
\left(\lambda-\tilde\zeta\cos^2\!\beta\right),
\label{eq:dcosbeta}\ee
where
\be
\tilde{\zeta}={\zeta'\cos^2\!\theta_\star \over  6\eta^{7/2}}.
\ee
Equation (\ref{eq:dcosbeta}) is the key result of this paper.
Note that while $\dot J_s$ depends on the (unspecified) spin-down
torque $|\cN_s|$, the evolution of the spin-orbit inclination angle
$\beta$ depends only on two dimensionless parameters:
$\lambda$ and $\tilde\zeta$. Although the precise values of these 
two parameters are uncertain (see Sect.~2), we expect $\lambda$ 
to lie in the range of  0.1 -- 1, and $\tilde\zeta$ to range from 
somewhat less unity to a few (for $\zeta'\simeq \zeta\sim 1$ and
$\eta\simeq 0.5$).

\begin{figure}
\begin{centering}
\vskip -2.4truecm
\includegraphics[width=9.3cm]{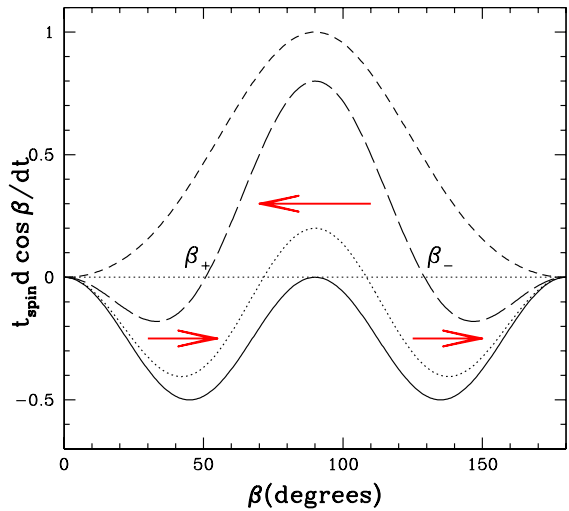}
\vskip -2.5truecm
\caption{The rate of change of the stellar inclination angle
$\beta$ for a fixed disc rotation axis 
(see Eq.~\ref{eq:dcosbeta}). From top to
bottom, the four curves correspond to $(\lambda,\tilde\zeta)=
(1,0.5),~(0.8,2),~(0.2,2)$ and $(0,2)$, respectively. 
The arrows indicate the direction of $\beta$ evolution. 
For $\tilde{\zeta}/\lambda<1$, the stellar
spin evolves towards alignment for all $\beta$ (see the short-dashed 
line); for $\tilde{\zeta}/\lambda>1$, the spin either evolves 
toward $\beta_+\neq 0$ or toward anti-alignment ($\beta=180^\circ$), 
depending on the initial values of $\beta$. The fiducial
spin evolution timescale 
$t_{\rm spin}$ is defined in Eq.~(\ref{eq:tspin}).}
\label{fig4}
\end{centering}
\end{figure}

Equation (\ref{eq:dcosbeta}) reveals the following behavior 
for the evolution of $\beta$ (see Fig.~\ref{fig4}):

(i) For $\lambda=0$, equation (\ref{eq:dcosbeta}) describes the effect 
of the magnetic warping torque acting alone on the star. 
This torque always pushes the stellar spin toward anti-alignment
with $\hat\bl$ (see Fig.~\ref{fig1}). In particular, the aligned
state ($\beta=0$) is unstable: For $\beta\ll 1$, we have
$\dot\beta/\beta=\tilde\zeta/t_{\rm spin}$. On the other hand, 
the perpendicular state ($\beta=\pi/2$) represents a ``bottleneck'' where
the warping torque $\bcN_w$ vanishes: For $\beta=\pi/2+\Delta$ with $|\Delta|\ll 1$,
we have $\dot\Delta/\Delta\simeq (\tilde\zeta/t_{\rm spin})\Delta$.
Thus, when the outer disc orientation is fixed (see Section 5), starting from
a small $\beta$, it would take infinite time to cross this $90^\circ$ barrier.

(ii) For $\tilde{\zeta}/\lambda<1$: Regardless of the initial
$\beta$, the spin always evolves towards alignment.

(iii) For $\tilde{\zeta}/\lambda>1$: There are two possible
directions of $\beta$ evolution, depending on the initial value of
$\beta$. The condition $d\cos\beta/dt=0$ yields
two ``equilibrium'' states ($\beta_+$ and $\beta_-$), given by
\be
\cos{\beta_{\pm}} = \pm \sqrt{\lambda/\tilde{\zeta}}.
\ee
Of the two equilibria, one is stable ($\beta_+$) and the other is
unstable. For $\beta(t=0) < \beta_-$, the system will 
evolve towards a misaligned prograde state $\beta_+$;
for $\beta(t=0) > \beta_-$, the system will evolve towards the
anti-aligned state ($\beta = 180^\circ$).

Note that the timescale to change the stellar spin (Eq.~[\ref{eq:tspin}])
can be written as $t_{\rm spin}=(J_s/M_\star l_{\rm in})
(M_\star/{\dot M})$, where $l_{\rm in}=\sqrt{GM_\star\rin}$ is the specific
angular momentum of the accreting gas at $\rin$. The disk lifetime 
(observed to be around 10~Myrs) is a few times less than $M_\star/\dot M$
if one uses the appropriately averaged value of $\dot M$. Since 
$J_s/(M_\star l_{\rm in})\ll 1$, this implies that $t_{\rm spin}$ 
is typically less than the disk lifetime, i.e., significant change of the 
stellar spin can be achieved during the disk lifetime. Of course, in the earlier
phases of the protostars (e.g., Class-0 T Tauri, with $\dot M
\sim 10^{-5} M_\odot/{\rm yr}$), $t_{\rm spin}$ is much shorter 
than 1~Myrs, while in the later phases $t_{\rm spin}$ is longer.

\section{Application to Spin-Orbit Misalignment in Exoplanetary Systems}

The results of previous sections clearly show that, contrary to
the standard assumption, the spin axis of a magnetic protostar does not 
necessarily align with the axis of its circumstellar disc. Therefore, 
a planet formed in the disc may have its orbital axis misaligned with
the stellar spin, even before any few-body interactions
(such as strong planet-planet scatterings and Kozai interactions)
take place. A clear prediction of the expected spin-orbit 
misalignment angle and its distribution for an ensemble of 
planetary systems is complicated by the fact that the physics 
of magnetic star -- disc interaction is complex: Even with our 
general parameterized model, the evolution of $\beta$ depends
on two unknown dimensionless parameters $(\lambda,\tilde\zeta)$.
The final misalignment angle and its distribution also depend on 
the usual parameters associated with protoplanetary discs
(such as the accretion rate and lifetime), as well as on the initial
conditions for the disc direction relative to the stellar spin.

We now discuss two scenarios to illustrate the possible
outcomes for the spin-orbit misalignment angles in exoplanetary systems.

\subsection{Scenario (a)}

Here we consider the case where the collapsing/accreting materials that 
form the protostar and its disc all have angular
momentum axes approximately in the same direction. This would lead to 
an initial state where the stellar spin and the disc axis
are approximately aligned (with small $\beta$). 
As we showed in Sect.~4, for systems 
with $\tilde\zeta/\lambda<1$, the angle $\beta$ will decrease and
the spin-orbit alignment will be enhanced. For systems with
$\tilde\zeta/\lambda>1$, however, the angle $\beta$ will increase
towards $\beta_+$ (see Fig.~\ref{fig4}).

In the idealized situation where the disc axis $\hat\bl$ is fixed
in time, we would expect that some planetary systems to have 
$\beta$ close to zero, while others to have $\beta$ clustered around
values somewhat less than $\beta_+$. Since $\beta_+=\cos^{-1}
\sqrt{\lambda/\tilde\zeta}$ are
different for different systems, we may not expect a strong clustering.
Note that because $\beta_+$ is always less than $90^\circ$, no planet 
in a retrograde orbit can be produced in this idealized situation. 
However, if we consider the possibility
that the axis of the outer disc changes in time
-- as might be expected because the gas
that feeds the outer disc does not necessarily have a fixed rotation axis
or because the outer disc is perturbed by a distant star in a cluster,
then planets on retrograde orbits may be produced.
To achieve this, the outer disc axis must vary with sufficient amplitude
[larger than $(\beta_--\beta_+)$] and on sufficiently short 
timescale (shorter than the spin evolution time), and the inner disc
axis must adjust (via bending wave propagation or viscous diffusion) 
quickly to the variation of the outer disc, so that a system at
the $\beta<\beta_+$ state may be pushed over to the $\beta>\beta_-$ state
--- We will show in Paper II that this is possible.

\subsection{Scenario (b)}

Here we consider the case where the initial stellar spin axis and the
disc axis are randomly distributed with respect to each other.
This might be expected if the turbulent gas in a molecular cloud
core that feeds into the disc falls in with random directions.
Suppose at the time of planetary formation,
the stellar spin is determined by the accumulative angular momentum 
accretion, then the stellar spin axis may be quite different from
the ``current'' disc axis (e.g., Bate et al.~2009), and the distribution
of the spin-orbit misalignment angle may be quite broad.
The important problem that we wish to address here is: How the
distribution function $f(\psi,t)$ evolves further in time?
(Here we define the angle between $\hat\bl$ and $\hat\bomega_s$
as $\psi=\beta$, in agreement with the notation used in observational
papers; see Triaud et al.~2010).

Without the magnetic warping effect discussed in Sect.~4,
the systems will evolve toward alignment, therefore an initial
random distribution, $f(\psi,0)=\sin\psi/2$, will become increasingly
more peaked towards $\psi<90^\circ$ as time passes.
However, when the magnetic warping effect in taken into account, 
a variety of outcomes become possible, as it is clear from our results
in Sect.~4.  

\begin{figure}
\begin{centering}
\includegraphics[width=8cm]{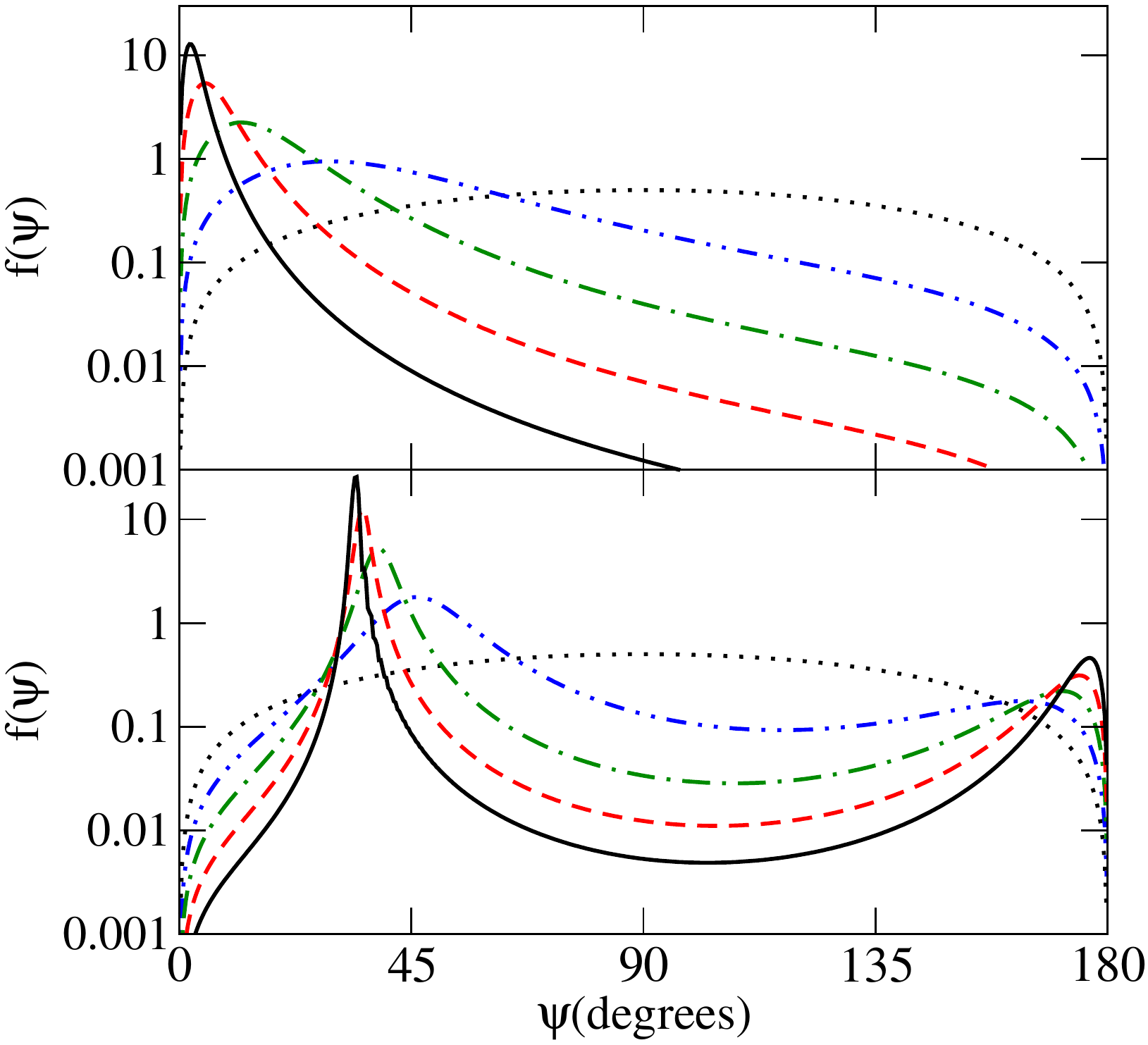}
\caption{Distribution of the angle $\psi$ between the stellar rotation axis
and the disc axis at $t/t_{\rm spin}=0$ (dotted line), $1,2,3,4$
  (solid line), for $\tilde{\zeta}=0.125$ (upper panel) 
  and $\tilde{\zeta}=\sqrt{2}$ (lower panel), both with $\lambda=1$. The initial
  distribution is isotropic.}
\label{fig5}
\end{centering}
\end{figure}

Consider a simple model where $\lambda,\,\tilde\zeta$ and 
$t_{\rm spin}$ are constant in time and the same for different systems. 
Then the evolution of $f(\psi,t)$ is governed by the equation
\be
{\partial\over\partial t}f(\psi,t)=-\frac{\partial}{\partial\psi}
\left[f(\psi,t) \frac{d}{dt}\psi\right], \ee 
where the time derivative of $\psi=\beta$ is given by 
Eq.~(\ref{eq:dcosbeta}). In Fig.~\ref{fig5}, we plot
$f(\psi,t)$ between $t=0$ and $t=4t_{\rm spin}$ for
$(\lambda,\tilde\zeta)=(1,0.125)$ and $(1,\sqrt{2})$, assuming an
initial isotropic distribution $f(\psi,0)=(\sin{\psi})/2$.  We see
that, as expected from the behavior depicted in Fig.~\ref{fig4}, for
$\tilde\zeta/\lambda<1$, the distribution function $f(\psi,t)$ will
become increasingly peaked at $\beta<90^\circ$.  For
$\tilde\zeta/\lambda>1$, however, $f(\psi,t)$ evolves into a
double-peaked function, with one of the peaks located 
around $\beta_+$ and the other close to $180^\circ$.

Obviously, to determine the true distribution $f(\psi,t)$
for an ensemble of planetary systems, one
must further ``average'' over the distribution of
$(\lambda,\tilde{\zeta})$ for different systems.
These parameters are largely unconstrained.
Nevertheless, we can reasonably expect to see a significant 
number of systems concentrating at small $\psi$ (all cases with
$\tilde{\zeta}/\lambda<1$), and the rest distributed between all possible
values of $\beta_+$ (which is less than $90^\circ$), 
or close to $\psi=180^\circ$. 

\section{Discussion}

We have shown in this paper that the angle between
the spin axis of a magnetic protostar and the axis of its disc
may have a wide range of values: some systems are expected to be
aligned, but some are expected to be misaligned. As a result, 
we would expect the spin-orbit inclination angle in exoplanetary 
systems to have a broad distribution, with some systems
aligned, while others highly misaligned, even without any 
additional/subsequent physical processes that may affect this angle.

While the results of this paper are compatible with the possibility
that planet-planet scatterings or Kozai interactions play a role in
determining the orbital characteristics of some close-in planets, they
nevertheless weaken the challenges posed against the disc-migration
scenario for the origin of most hot Jupiters. 
Most likely, both few-body interactions and disc migration are needed
to explain the observed period distribution of exoplanetary systems.

Currently, there is no measurement of the spin-disc misalignment angle
in any accreting T Tauri star systems, although in principle this angle may
be constrained from spectropolarimetry observations (e.g., Hussain et
al. 2009; Donati et al.~2010) and detailed modeling of the
variabilities of T Tauri stars. 
Alternatively, spin-disc alignment may be tested by measuring the
$v\sin i_\star$ and the rotation period of the star, and comparing the
resulting stellar inclination $i_\star$ with the disc inclination
$i_{\rm disc}$, which may be constrained from the jet direction (for T
Tauri discs) or the sky-projected shape of debris
discs\footnote{Recently, Watson et al.~(2010) carried out such an
analysis for several debris disc systems and found no significant 
difference between $\sin i_\star$ and $\sin i_{\rm disc}$. Note that 
$i_\star=i_{\rm disc}$ does not necessarily imply alignment between
the spin axis and the disc axis. Also, systematic uncertainties in
estimating $i_{\rm disc}$ need to be taken into account. For example,
for HD 22049 (one of the best cases studied by Watson et al), 
the disc inclination is consistent with face-on ($i_{\rm disc}\lo 25^\circ$;
K. Stapelfeldt 2010, private communication; see Backman et al.~2009).}.
On the theoretical side, the spin-disc
misalignment angle is determined by the uncertain values of two
dimensionless parameters, $\lambda$ and $\tilde\zeta$ (see Section 4)
and their possible dependence on various unknown quantities (such as
the mass accretion rate), as well as on the initial condition (see Section 5)
and the age of the disc-accretion phase. However, these two parameters may
potentially be calibrated from observations of a large sample
of T Tauri stars. With these input parameters, we can
then compare theoretical predictions (which include not only the 
process studied in this paper but also few-body interactions that
may change the spin-orbit angles) with the observed distribution of
misalignment angles between planets' orbits and the spins of their
mature host stars (Triaud et al.~2010).

A more direct observational test is to measure the spin-orbit 
angle for multiple planets systems. Few-body interactions would 
produce different angles for different planets in the same system.
In contrast, the magnetosphere -- disc interaction studied 
in this paper would lead to 
the reorientation of the stellar spin while preserving the orbital
plane of the planetary systems.  Up to now, there is no known 
multiple-planet systems with measured stellar spin and planet 
orbital angular momentum vector.  In the context of the solar 
system, the orbits of most planets, including those of  
Jupiter and Saturn, are confined within $\sim 2^\circ$ from the ecliptic.
Although the Sun's spin vector is misaligned by $7^\circ$ 
with the vector normal to the ecliptic plane of the solar system,
this modest difference may or may not provide adequate support for 
the magnetosphere -- disc interaction scenario.

According to the scattering/Kozai scenario, planets venture into the
proximity of their host stars with a diverse degree of spin-orbit
misalignment.  Yet, the misaligned systems tend to have parent stars
with mass greater than $1.2M_\odot$ (Schlaufman 2010; Winn et
al.~2010).  Based on this scenario, Winn et al. (2010) suggested that
spin-orbit misalignment for planets around solar type stars has been
damped out by tidal dissipation in the star, whereas such dissipation
may be less efficient for more-massive stars (with $M>1.2M_\odot$)
because of their shallow or non-existent outer convection zones (e.g.,
Barker \& Ogilvie 2009). Nevertheless, the tidal dissipation time
scale is a rapidly increasing function of planets' orbital semi-major
axis.  Therefore, according to the tidal interaction hypothesis, the
spin-orbit misalignment may nonetheless be preserved for planets
around solar type stars with periods longer than a few days.

In the scenario that the spin-orbit misalignment is induced by 
the magnetosphere -- disc interaction, tidal damping of the misalignment 
angle can still operate more effectively around solar type stars than 
more massive stars.  In addition, the magnitudes of $\lambda$ and 
$\tilde\zeta$ and therefore the resultant value of $\beta$ are
functions of the stellar mass.  In this case, the correlation between
spin-orbit misalignment and the stellar mass would be independent of 
the planets' orbital period (for periods longer than a few days). 
The confinement of multiple planets
in a common orbital plane would also be preserved.  These observational
tests, though challenging, will eventually provide clues and constraints
on the origin of hot Jupiters as well as that of the spin -- orbit 
misalignment.

\section*{Acknowledgments}

We thank Dan Fabrycky and other participants of the KITP Exoplanet
program (Spring 2010) for useful discussions.  D. Lai and D. Lin
acknowledge the hospitality of the Kavli Institute for Theoretical
Physics at UCSB (funded by the NSF through Grant PHY05-51164), where
this work started. D. Lai also thanks KIAA (Beijing) for hospitality.
This work has been supported in part by NASA Grant NNX07AG81G,
NNX07A-L13G, NNX07AI88G, NNX08AL41G, NNX08AM84G, and NSF grants 
AST-0707628, AST-0908807 and AST-1008245.



\end{document}